\documentclass[prl,twocolumn,graphicx,amssymb,floatfix]{revtex4}
\usepackage{graphicx}
\usepackage{color}
\usepackage{soul}
\usepackage{gensymb}
\begin{document}

\title{Comment on ``Non-representative Quantum Mechanical Weak Values"}
\author{Alon Ben Israel and L. Vaidman}
\affiliation{ Raymond and Beverly Sackler School of Physics and Astronomy\\
 Tel-Aviv University, Tel-Aviv 69978, Israel}

\begin{abstract}
Svensson [Found.  Phys. \textbf{45}, 1645 (2015)] argued that the concept of the weak value of an observable of a pre- and post-selected quantum system cannot be applied when the expectation value of the observable in the initial state vanishes. Svensson's argument is analyzed and shown to be inconsistent using several examples.
\end{abstract}

\maketitle

In a paper with a provocative title ``Non-representative quantum mechanical weak values'' \cite{Svensson15} Svensson claims that under certain conditions the weak value does not represent an undisturbed quantum system. Weak values were introduced as outcomes of weak measurements in the limit of weak coupling \cite{AAV}, but Svensson argues that in some cases the weak limit is discontinuous and, therefore, we cannot associate the weak value with the undisturbed system as it is commonly done. Here we refute Svensson's arguments.

Svensson begins by outlying a procedure to arrive at (the real part of) the weak value of an observable of a quantum system. This scheme entails taking the coupling constant $g$ to the limit $g\rightarrow0$. This approach is different from the one originally used in the introduction of weak values, where the limit taken, $\Delta\rightarrow\infty$, was related to the quantum state of the measuring device \cite{AAV}. The two cases are not entirely equivalent as $g$ is a property of the interaction and $\Delta$ is a property of the pointer, and so it is not clear that Svensson's argument can be stated for the original proposal. However, if we wish to consider discrete pointers (such as the interaction of a spin with another spin) we must indeed use the limit $g\rightarrow0$, and so it is worth considering Svensson's critique.

\section{``Discontinuity'' of weak value}

Svensson claims that the outlined scheme is `discontinuous', in that taking ${g\rightarrow0}$ does not represent the undisturbed $g=0$ case if $\langle in|{\cal S}|in\rangle=0$. That is, the process is discontinuous if the operator whose weak value we wish to measure takes the system's initial state (the pre-selected state) to an orthogonal one. However, no discontinuity is demonstrated. While ${\cal S}|in\rangle=0$ may be orthogonal to $|in\rangle$, nowhere does this suggest a discontinuity when taking $g\rightarrow0$. The operator describing the evolution (their Eq. (2)) is $exp(-ig{\cal S}\otimes P_M)$, and to the first order in $g$ the state of the system and the pointer following the interaction is, as Svensson shows,
 \begin{equation}\label{U}
 |in\rangle \otimes |m\rangle - ig{\cal S}|in\rangle \otimes P_M|m\rangle.
  \end{equation}
  It is very clear from this expression that the total state does in fact merge smoothly with the state in the $g=0$ case when taking $g\rightarrow0$.

\section{ Svensson's special condition}

  Svensson's special condition for which he argues that the weak value ``has a problem'' is $\langle in|{\cal S}|in\rangle=0$.
 He focuses on the second term of (\ref{U}) alone, writing that this is ``the source of the weak value''. He does not explain what it might mean: the weak value depends on several factors. Anyway,  in the weak limit, the total state of the system is nearly unchanged whatever the effect of ${\cal S}$ on the system may be. The `derailing' of the system is always very small compared to the original state which remains unchanged, and it makes no difference whether this small addition is orthogonal to the original state or not.  
 \begin{sloppypar}
 Svensson then continues with a few examples where we might have ${\langle in|{\cal S}|in\rangle=0}$. He reiterates that this is ``a problem'' but does not  demonstrate what this problem is, nor how usage of standard two-state vector formalism  in these cases causes any sort of unwanted or inconsistent result. We argue that no problem exists.
 \end{sloppypar}
We will consider now several examples in which Svensson's criterion runs into problems.
First, consider a system with a pre-selected state which is an eigenstate of  ${\cal S}$, with eigenvalue zero. Here, the measurement interaction does nothing at all. It is clear that the resulting weak value of zero is in fact descriptive of the fact that nothing happened. There is no derailment. Svensson addresses the case of ${\cal S}|in\rangle = 0$, but does not consider this simple example. If one accepts that the example given is clearly descriptive (there is no question that the pointer state does not change at all), then the condition $\langle in|{\cal S}|in\rangle=0$ is not relevant in this case and is inconsistent. We will consider less trivial examples in the following sections.

\section{ Svensson's  condition for  a spin-$\frac{1}{2}$ particle example}
\begin{sloppypar}
Consider next  a spin-$\frac{1}{2}$ particle, pre-selected in ${|in\rangle = |{\uparrow}_x\rangle}$ and postselected in $|out\rangle = |{\uparrow}_z\rangle$ with an operator ${\cal S}=\sigma_z$. In this case ${\cal S}|in\rangle$ is nonzero, but $\langle in|{\cal S}|in\rangle=0$. According to Svensson, this case should be problematic. However, let us now, instead of measuring the spin in the ${\hat z}$ direction, perform a measurement of spin in the ${\hat z}+{\hat x}$ direction, ${{\cal S}_+ \equiv \frac{1}{\sqrt{2}}\left(\sigma_z+\sigma_x\right)}$ and a measurement of spin in the ${\hat z}-{\hat x}$ direction, ${{\cal S}_- \equiv \frac{1}{\sqrt{2}}\left(\sigma_z-\sigma_x\right)}$. Since ${{\cal S} = \frac{1}{\sqrt{2}}\left({\cal S}_+ + {\cal S}_-\right)}$, and since the weak values are additive, ${(A+B)_w=A_w+B_w}$, we can measure weak value of ${\cal S}$ as the sum of weak values of ${\cal S}_+$ and ${\cal S}_-$. These weak measurements should not have a problem since $\langle in|{\cal S}_+|in\rangle=\frac{1}{\sqrt{2}}$ and $\langle in|{\cal S}_-|in\rangle=-\frac{1}{\sqrt{2}}$, i.e. none of these expressions vanish.
  By Svensson's arguments, the two resulting weak values are descriptive of the system but their sum is not.
\end{sloppypar}
  Now, imagine that for the spin measurement of $\sigma_z$ the pre- and post-selection are flipped, meaning a pre-selection of $ |{\uparrow}_z\rangle$ and post-selection of $ |{\uparrow}_x\rangle$. According to Svensson,  there is no problem in this case, $\langle in|{\cal S}|in\rangle=1$ and the system is ``well-behaved''. However the two-state vector formalism is time-symmetric. It makes little sense, then, to base an objection on the effect of $S$ on $|in\rangle$ and not to address its effect on $|out\rangle$. His objection is therefore inconsistent or, at best, incomplete.

  \section{ Nested interferometer example}
  
  Svensson continues with analysis of the example of the nested Mach-Zehnder interferometer (based on \cite{past}). His analysis is inconsistent with his basic claim that considering the weak value of ${\cal S}$ is problematic when ${\langle in|{\cal S}|in\rangle=0}$. In the nested interferometer ${\langle in|\Pi_B|in\rangle\neq0}$ and $\langle in|\Pi_C|in\rangle\neq0$. Thus, the weak values $(\Pi_B)_w$ and $(\Pi_C)_w$ are not supposed to be problematic, but their nonvanishing values are the main story of the controversy of the nested interferometer.

Neither the interaction in $B$  nor the interaction in $C$   derail the system into an orthogonal state. Even if we accept Svensson's dismissal of the $D_3$ component which does not reach the detector, we should not ignore the $A$-arm component since it does reach the final detector. Svensson focuses only on the $E$-arm component, claiming that there is a problem since it is orthogonal to the undisturbed system. However, one cannot simply look at only part of the evolved state and claim that the state is orthogonal.

Note that in the nested interferometer we do have $\langle in|\Pi_E|in\rangle=0$, which fits Svensson's approach for a problematic point, but the absence of the particle in $D$ is as strange as the absence of the particle in $E$, while $\langle in|\Pi_D|in\rangle\neq0$.

We admit that the points in arms $E$  and $D$ are different from some arbitrary points outside the interferometer. Nonzero $g$ does not lead to a nonvanishing probability of finding the particle outside the interferometer. This is one of the reasons why arms $E$  and $D$ got the status of ``secondary presence'' in \cite{secondary}. However, the arms $E$  and $D$ cannot get the status of primary presence which  arm $A$ has, since in the weak limit the magnitude of the trace in $A$ is infinitely larger than the magnitude of the trace in $E$ and $D$. In contrast, the presence in $B$  and $C$ has primary status, because the trace there is of the same order as in arm $A$.

\section{Concluding remarks}

   Svensson's conclusions are puzzling: ``the question whether the weak value generating state $g {\cal S}|in\rangle$ lies entirely in an orthogonal subspace or not can be given a yes or no answer for all nonvanishing values of $g$. And if the answer is yes, then this state does not represent ${\cal S}$, implying that neither does the weak value: it is constructed from a state that is orthogonal to the state of the undisturbed system and does not describe ${\cal S}$." Nowhere does Svensson define what `describing' or `representing' a state or an operator means.  The statement itself is furthermore perplexing: a weak value is not a state, nor is it `constructed' from one.

In summary, we have shown that Svensson's claim about discontinuity of weak value does not hold. His arguments are not based on rigorous definitions, and we have shown that his main criterion contradicts time symmetry of the situations for which weak values are defined and runs into inconsistency in several examples.

 We thank Lukas Knips, Jan Dziewior and Jasmin Meinecke for useful discussions. This work has been supported in part by the  Israel Science Foundation  Grant No. 1311/14  and the German-Israeli Foundation for Scientific Research and Development Grant No. I-1275-303.14.

\end{document}